\documentclass[12pt,letterpaper]{article}

\usepackage{amsmath,amssymb,calc}
\usepackage{graphicx}

\newcommand\be{\begin{equation}}
\newcommand\ee{\end{equation}}
\newcommand{\bea}{\begin{eqnarray}}
\newcommand{\eea}{\end{eqnarray}}

\newcommand{\nn}{\nonumber}
\newcommand{\pd}{\partial}

\def\id{\protect{{1 \kern-.28em {\rm l}}}}

\def\id{\protect{{1 \kern-.28em {\rm l}}}}

\begin{document}

\begin{titlepage}
\begin{center}
\hfill \\
\vspace{2cm}
{\Large {\bf A Gravity Dual of Ultra-slow Roll Inflation
\\[3mm] }}

\vskip 10mm

{\bf Lilia Anguelova\\
\vskip 0.5cm  {\it Institute for Nuclear Research and Nuclear Energy}\\
{\it Bulgarian Academy of Sciences, Sofia 1784, Bulgaria}\\
{\tt anguelova@inrne.bas.bg}}

\vskip 6mm

\end{center}

\vskip .1in
\vspace{1cm}

\begin{center} {\bf Abstract}\end{center}

\vspace{-1cm}

\begin{quotation}\noindent

We study time-dependent deformations of a certain class of backgrounds in type IIB supergravity. These backgrounds are solutions of a five-dimensional consistent truncation, relevant for gauge/gravity duality, which have the form of $dS_4$ foliations over a fifth (radial) direction. We investigate time-dependent deformations of those solutions in the search for gravitational duals of models of glueball inflation. A particular starting ansatz enables us to find a class of analytical solutions, corresponding to an ultra-slow roll inflationary regime. This regime may play a role in understanding the low $l$ anomaly in the power spectrum of the CMB.

\end{quotation}

\end{titlepage}

\eject

\tableofcontents

\section{Introduction}

The observed large-scale homogeneity and isotropy of the present-day Universe can be explained by the existence of a period of inflationary expansion in the Early Universe. According to the standard field-theoretic lore, such an expansion is driven by the potential energy of a fundamental scalar field called inflaton. In addition, certain ratios of the inflaton potential and its derivatives have to satisfy so called slow roll conditions for the duration of the inflationary stage. The reason is to ensure a nearly-scale invariant spectrum of perturbations, as well as a large enough number of e-folds. However, it is well-known \cite{CLLSW,DRT} that one of those conditions tends to be violated by quantum corrections. This is because the relevant ratio, denoted by $\eta$, is proportional to the second derivative of the potential and, thus, to the mass of the inflaton. Hence, quantum corrections to the inflaton mass lead to a premature exit from the inflationary stage. This is known as the $\eta$-problem in inflationary model building. 

There is a huge literature devoted to the search for resolutions to this problem. A large amount of it is on studying models that have potentials with a shift symmetry, most notably axion monodromy inflation ones \cite{AxionMonInfl}. Another class of models is based on having a composite inflaton \cite{CJS,BCJS}. The idea is that if the inflaton is a composite state in a strongly-coupled gauge sector, then its mass is dynamically fixed and so there is no $\eta$-problem. However, strongly-coupled gauge dynamics is, clearly, beyond standard perturbative QFT methods. This gives us motivation to study composite-inflation model building by using a relatively new method originating from string theory, namely the gauge/gravity duality \cite{JM,GKP,EW,IMSY}. The latter is a powerful technical tool for investigating the non-perturbative regime of gauge theories via dual gravitational backgrounds.

As an initial step in this program, in \cite{ASW, ASW2} we found a certain class of gravity duals whose noncompact 5d space has the form of $dS_4$ fibered over a radial direction. Here we will continue this investigation by looking for deformations around the $dS_4$ background, that represent a 4d spacetime with time-dependent Hubble parameter. Such deformed solutions would provide gravity duals of composite inflation models, in which the inflaton is a glueball. We should note that similar ideas were explored in \cite{AB}, which studied deformations of a number of prominent gravity duals of gauge theories living in 4d Minkowski spacetime.\footnote{See also \cite{EFK} for another holographic model, in which the inflaton is a quark condensate.} The noncompact part of the relevant 10d solutions was deformed there to having a 4d de Sitter component instead of a Minkowski one. The main conceptual difference between these works and our considerations is that in \cite{AB} the inflaton arises from the open string sector, as the position of a D3-brane probe, whereas in our case its origin is in the closed string sector, as a state in the spectrum of the (scalar) background fluctuations that describe glueballs in the dual gauge theory.

The gravity duals we study are solutions of a 5d consistent truncation of type IIB supergravity, that was established in \cite{BHM}. This effective theory provides a useful unifying framework for gauge/gravity duality considerations, since it contains as special solutions a large variety of famous backgrounds like the Maldacena-Nunez \cite{MN} and Klebanov-Strassler \cite{KS} duals of ${\cal N}=1$ SYM, as well as their more recent deformations \cite{NPP,ENP,EGNP} that are dual to gauge theories with multi-scale dynamics. In \cite{ASW}, we found several classes of solutions to the relevant 5d equations of motion, which have a metric with $dS_4$ slicing. This is of use for studying strongly-coupled gauge theories in $dS_4$ space. However, here we are interested in 5d solutions with inflating 4d spacetime and time-dependent 5d scalars (i.e., inflatons). Since the main cosmological observables, namely the scalar spectral index $n_s$ and the tensor-to-scalar ratio $r$, are entirely determined by the Hubble parameter and inflaton as functions of time, those solutions would provide models of glueball inflation. 

To build such models, we study time-dependent deformations of the solutions of \cite{ASW}. For a particular initial ansatz, we find a gravitational dual, which turns out to describe an ultra-slow roll regime. This regime is unstable and can only last a few e-folds, before transitioning to standard slow-roll. Thus, this is not a complete description of cosmological inflation by itself. Nevertheless, the presence of such a transient stage at the time, when the largest CMB scales we observe today exited the horizon, could play an important role in explaining the anomaly \cite{CMBanomaly} in the CMB power spectrum at low values of the multipole moment $l$. 

This paper is organized as follows. In Section 2, we review the 5d consistent truncation of \cite{BHM}, as well as the analytical solution of \cite{ASW}. We also introduce the ansatz we will use to look for deformations around the latter. In Section 3, we find a solution to the field equations for the relevant 5d scalars and metric functions. In Section 4, we analyze the implications of our solution for glueball inflation and show that it gives an ultra slow-roll model. In Section 5, we summarize the present work and discuss perspectives for building duals of standard slow-roll inflation. Finally, in the Appendix we give technical details about our inflaton solution at the next subleading order in a certain expansion in terms of a small parameter.

\section{5d theory and deformation ansatz}
\setcounter{equation}{0}

In this section we will describe the 5d consistent truncation of type IIB supergravity, whose equations of motion we are interested in. We will also recall the analytical solution with $dS_4$ slicing that was found in \cite{ASW}. Finally, we will introduce our ansatz for small deformations around that solution.

\subsection{Effective 5d theory}

Our goal will be to find a particular kind of solutions to the equations of motion of the effective 5d theory provided by the consistent truncation of \cite{BHM}. So let us begin by briefly reviewing relevant material about the latter.

The bosonic field content of type IIB supergravity is the following: 10d metric,  string dilaton, RR scalar, NS and RR 3-form fields and RR 5-form field. Substituting a certain ansatz for those fields into the 10d IIB action and integrating out five compact dimensions, one is left with an effective 5d action of the form \cite{BHM}:
\be \label{5daction}
S = \int d^5 x \sqrt{- det g} \left[ - \frac{R}{4} + \frac{1}{2} G_{ij} (\Phi) \pd_I \Phi^i \pd^I \Phi^j + V (\Phi) \right] \, ,
\ee
where $R$ is the Ricci scalar of the 5d metric $g_{IJ}$, while $\{\Phi^i\}$ is a set of 5d scalars arising from various components of the 10d fields, including warp factors, $G_{ij} (\Phi)$ is a diagonal sigma-model metric and $V(\Phi)$ is a complicated potential; for more details see \cite{BHM} as well as the summary in \cite{ASW}. We will concentrate here on the same subtruncation as in \cite{ASW}, meaning that we will take the 10d NS 3-form to vanish. Hence the set of 5d scalars is:
\be 
\{\Phi^i (x^I)\} = \{ p (x^I), x (x^I), g (x^I), \phi (x^I), a(x^I), b(x^I) \} \,\, , 
\ee
where we have used the same notation as in \cite{BHM,ASW}. Then the nonvanishing components of the metric $G_{ij}$ are:
\be \label{SigmaMM}
G_{pp} = 6 \,\,\, , \,\,\, G_{xx} = 1 \,\,\, , \,\,\, G_{gg} = \frac{1}{2} \,\,\, , \,\,\, G_{\phi \phi} = \frac{1}{4} \,\,\, , \,\,\, G_{aa} = \frac{e^{-2g}}{2} \,\,\, , \,\,\, G_{bb} = \frac{P^2 e^{\phi - 2x}}{2} \,\,\, ,
\ee
where $P = const$ is the coefficient of the RR 3-form flux. Finally, the equations of motion that follow from the action (\ref{5daction}) are:
\bea \label{EoM}
\nabla^2 \Phi^i + {\cal G}^i{}_{jk} \,g^{I J} (\pd_I \Phi^j) (\pd_J \Phi^k) -V^i &=& 0 \quad , \nn \\
- R_{IJ} + 2 \,G_{ij} \,(\pd_I \Phi^i) (\pd_J \Phi^j) + \frac{4}{3} \,g_{IJ} V &=& 0 \quad ,
\eea
where $V^i = G^{ij} V_j$ \,, \,$V_i = \frac{\pd V}{\pd \Phi^i}$ \,and ${\cal G}^i{}_{jk}$ are the Christoffel symbols of the metric $G_{ij}$\,.

A wide variety of solutions of the system (\ref{EoM}) can be identified, in the context of the gauge/gravity duality, with gravitational duals of strongly coupled gauge theories. More precisely, one can find duals of (some) 4d gauge theories living in Minkowski space by solving (\ref{EoM}) with the metric ansatz:
\be \label{flatM}
ds_5^2 = e^{2A(z)} \eta_{\mu \nu} dx^{\mu} dx^{\nu} + dz^2 \,\, .
\ee
In this context, the scalar fields $\Phi^i$ describe glueball states in the dual gauge theory. For example, the Maldacena-Nunez solution \cite{MN} is obtained for 
\be
Q=0 \quad , \quad b=a \quad , \quad x=\frac{1}{2} g - 3p \quad , \quad \phi = -6p - g - 2 \ln P
\ee
and some particular functions $a=a(z)$, $g=g(z)$ and $p=p(z)$; for more details see \cite{BHM,PT}.

In the following we will investigate solutions, in which the ansatz (\ref{flatM}) is modified so that the 4d flat metric $\eta_{\mu \nu}$ is substituted by that of an expanding spacetime. The starting point for our considerations will be an analytical solution with a 4d de Sitter metric, instead of the Minkowski one, that was obtained in \cite{ASW}. It was discussed there that this solution has a lot of similarities with the ${\cal N} = 1$ gravitational dual of \cite{NPP}, corresponding to a certain walking gauge theory. Understanding better the relation between the two could be of great importance for finding microscopic realizations of the solution of \cite{ASW} as some kind of a nonsupersymmetric deformation of a type IIB D-brane configuration. We hope to come back to this issue in the future. 

\subsection{Analytical solution}

In \cite{ASW}, we found solutions of the system (\ref{EoM}), which have a 5d metric of the form:
\be \label{metricans}
ds_5^2 = e^{2A(z)} \left[ -dt^2 + s(t)^2 d\vec{x}^2 \right] + dz^2 \,\, ,
\ee
where $s(t) = e^{{\cal H}t}$ with ${\cal H} = const$. In those solutions three of the six scalars $\Phi^i$ are identically zero, namely:
\be \label{gabzero}
g (x^I) = 0 \quad , \quad a (x^I) = 0 \quad , \quad b (x^I) = 0 \quad ,
\ee
whereas (some of) the remaining three scalars are nontrivial functions of $z$ only. More precisely, we found two four-parameter families of numerical solutions, for which all three scalars $p$, $x$ and $\phi$ have nontrivial $z$-profiles, and one three-parameter family of analytical solutions, for which only $p$ and $x$ do while $\phi$ vanishes identically. 

All of these solutions have a constant 4d Hubble parameter ${\cal H}$, as mentioned above, and thus are relevant for studying strongly-coupled gauge theories in de Sitter space. Here, however, we are interested in solutions with a time-dependent Hubble parameter. Recall that the latter is defined as:
\be \label{Hdef}
{\cal H} \equiv \frac{\dot{s}}{s} \,\,\, ,
\ee
where $\dot{s} \equiv ds / dt$. Since one of the inflationary slow roll conditions is \cite{DB}:
\be \label{SRcond}
- \frac{\dot{{\cal H}}}{{\cal H}^2} \,<\!\!< \,1 \,\,\,,
\ee
we can look for solutions with $\dot{{\cal H}} \neq 0$ by studying small deformations, in the sense that (\ref{SRcond}) is satisfied, around the constant-${\cal H}$ solutions of \cite{ASW}. To allow for such metric deformations, we will need to turn on time-dependence in (at least some of) the scalar fields. Hence the coupled system of field equations that we will need to solve will be rather involved. So, for simplicity and more insight, we will study here deformations around the analytical solution of \cite{ASW}. 

Let us recall the form of this solution. First, note that in the limit, in which it was derived, the 5d scalar potential reduces to (see (4.10) of \cite{ASW}):
\be \label{Vzero}
V = - N^2 e^{2 p - 2 x} \,\,\, ,
\ee
where $N = const$. Now, using the subscript $0$ to denote the metric and scalar functions pertaining to this specific solution, we have (see Section 4.1.3 of \cite{ASW}):
\bea \label{ZeroOrSol}
A_0 (z) &=& \ln (z+C) + \frac{1}{2} \,\ln \!\left( \frac{7}{3} {\cal H}_0^2 \right) \,\, , \nn \\
p_0 (z) &=& - \frac{1}{7} \ln (z+C) - \frac{1}{14} \,\ln \!\left( \frac{7 N^2}{9} \right) \,\, , \nn \\
x_0 (z) &=& - 6 \,p_0 (z) \qquad , \qquad \phi_0 = 0 \quad ,
\eea
where $C$ is an integration constant and the Hubble constant ${\cal H}_0$ is related to the 4d cosmological constant $\Lambda$, used in (4.30) of \cite{ASW}, via $\Lambda = 3 \,{\cal H}_0^2$. As discussed in \cite{ASW,ASW2}, the large $z$ asymptotics of this solution is consistent with ALD (asymptotically linear dilaton) behavior, despite superficial appearances. Note that the walking solution of \cite{NPP} has the same kind of asymptotics, which is one out of a number of similarities between the two solutions as pointed out in \cite{ASW}.  

\subsection{Deformation ansatz}

Let us now introduce our ansatz for small deformations around the zeroth order solution given by (\ref{ZeroOrSol}). For simplicity, we would like to turn on time-dependence in as few scalars as possible. In that regard, notice that the zeroth order solution for $\phi$ vanishes and also the scalar $\phi$ does not enter the potential (\ref{Vzero}) and thus $V^{\phi} = 0$. Therefore, it is convenient to choose $\phi$ to be the field that acquires time-dependence due to the deformations (i.e., to be the inflaton), while keeping the scalars $p$ and $x$ exactly the same as at zeroth order. One should keep in mind, though, that a time-dependent deformation might, in principle, have to be accompanied by a certain deformation of the $z$-dependence as well.\footnote{See, for example, \cite{EGM} and references therein.} So we will look for solutions, in which the scalar fields have the form: 
\be \label{TimeDepScalarAnz}
\phi = \phi (t,z) \quad , \quad p = p_0 (z) \quad , \quad x = x_0 (z) \,\,\, ,
\ee
whereas the 5d metric has the form:
\be \label{TimeDepMetricAnz}
ds_5^2 = e^{2A(t,z)} \left[ -dt^2 + e^{2H(t,z)} d\vec{x}^2 \right] + dz^2 \,\,\, .
\ee 

Since we are only interested in small deformations around (\ref{ZeroOrSol}), let us introduce a parameter $\gamma$, such that 
\be
\gamma <\!\!< 1 \,\, ,
\ee
and search for a solution as an expansion in powers of $\gamma$. Namely, we will make the following ansatz for the deformed scalar and metric functions:
\bea \label{Expansions}
\phi(t,z) &=& \gamma \,\phi_{(1)} (t,z) + \gamma^3 \phi_{(3)} (t,z) + {\cal O} (\gamma^5) \,\,\, , \nn \\
A(t,z) &=& A_{(0)}(z) + \gamma^2 A_{(2)}(t,z) + {\cal O} (\gamma^4) \,\,\, , \nn \\
H(t,z) &=& H_{(0)} (t) + \gamma^2 H_{(2)} (t,z) +{\cal O} (\gamma^4) \,\,\, ,
\eea
where 
\be \label{A0H0}
A_{(0)} (z) = A_0 (z) \qquad {\rm and} \qquad H_{(0)} (t) = {\cal H}_0 \,t \,\,\, .
\ee
For convenience, in the following we will denote:
\be
{\cal H}_0 \equiv h \,\,\, .
\ee 
Let us underline again that for the remaining two scalars we take the zeroth order expressions: $p (t,z) = p_0 (z)$ and $x (t,z) = x_0 (z)$. Note that the different powers of $\gamma$ in the $\phi$-expansion in (\ref{Expansions}), compared to the expansions of the warp factors $A(t,z)$ and $H(t,z)$, will play an important role in solving the system, as will become clear below. 

Finally, let us comment on the Hubble parameter of the deformed solutions we are looking for. From (\ref{Expansions})-(\ref{A0H0}), it is clear that the zeroth order value $h$ can be obtained as $h = \dot{H}(t,z)\big|_{\gamma = 0}$\,, where again we have denoted $\dot{}\equiv \frac{\pd}{\pd t}$ for convenience. Now, if the warp factor $A$ does not depend on $t$, then it is easy to realize that differentiating $H$ with respect to time provides the correct generalization for non-vanishing $\gamma$ as well. To see this, note that, in the metric ansatz (\ref{TimeDepMetricAnz}), the role of the scale factor $s(t)$ of (\ref{metricans}) is played by $e^{H(t,z)}$. Therefore, (\ref{Hdef}) generalizes to the following expression for the Hubble parameter:
\be \label{Hubpar}
{\cal H} \equiv \frac{1}{e^{H(t,z)}} \frac{\pd (e^{H(t,z)})}{\pd t} = \dot{H}(t,z) \,\, .
\ee
If it turns out that, at order $\gamma^2$ or higher, the warp factor $A$ becomes a nontrivial function of time, then one should first perform a coordinate transformation $t \rightarrow \tau$, absorbing this time dependence into the definition of $\tau$. Then, one can again apply (\ref{Hubpar}) with $t$ substituted by $\tau$. This discussion will be very useful in a later section.

\section{Solving the equations of motion}
\setcounter{equation}{0}

In this section we will look for solutions of the system (\ref{EoM}) with the ansatz (\ref{Expansions}) order by order in powers of the small parameter $\gamma$. The different powers of $\gamma$ in the expansion of $\phi$, compared to the expansions of the metric functions, will allow us to decouple the leading order equation of motion for $\phi$ from those for the warp factors. This technical point will enable us to solve the coupled system analytically.

\subsection{Scalar field equation}

Let us begin by computing the Christoffel symbols for the metric (\ref{TimeDepMetricAnz}). The nonvanishing components are:
\bea \label{Chris_symb}
&&\Gamma^t_{tt} = \dot{A} \quad , \quad \Gamma^t_{tz} = A' \quad , \quad \Gamma^t_{x^i x^i} = e^{2H} \left( \dot{A} + \dot{H} \right) \quad , \quad \Gamma^{x^i}_{x^i t} = \dot{A} + \dot{H} \,\, , \nn \\
&&\Gamma^{x^i}_{x^i z} = A' + H' \quad , \quad \Gamma^z_{tt} = A' e^{2A} \quad , \quad \Gamma^z_{x^i x^i} = - (A' + H') \,e^{2A+2H} \,\, ,
\eea
where $i = 1,2,3$ and $' \equiv \frac{\pd}{\pd z}$. Now, substituting (\ref{Expansions}) into $\nabla^2 \phi = \left( \pd_I \pd^I + \Gamma^I_{IJ} \pd^J \right) \phi$, we see that the leading order contribution, i.e. the one at order $\gamma$, is obtained from the zeroth order (i.e., $\gamma^0$) terms in $\nabla^2$ and thus the zeroth order Christoffel symbols. This is because $\phi$ itself is already of order $\gamma$ or higher. Recall also that $V^{\phi} = 0$, since the potential (\ref{Vzero}) is independent of $\phi$. So, at leading order, the $\phi$ field equation becomes $\nabla^2_{(0)} \phi_{(1)} = 0$, or in more detail:
\be \label{phi1}
\ddot{\phi}_{(1)} + 3 \,h \,\dot{\phi}_{(1)} = e^{2 A_{(0)}} \!\left( \phi''_{(1)} + 4 A'_{(0)} \phi'_{(1)} \right) \,\, .
\ee
At the next, $\gamma^3$, order we will find an equation for $\phi_{(3)}$, that also includes mixing between $\phi_{(1)}$ (or its derivatives) and $A_{(2)}$ or $H_{(2)}$ (or their derivatives); for more details see the Appendix. However, as we will show below, finding $\phi_{(1)}$ allows one to completely solve for the leading order metric deformations $A_{(2)}$ and $H_{(2)}$. Since this is enough for our purposes, we will not discuss here the field equation for $\phi_{(3)}$ any further.\footnote{Note, though, that knowing $\phi_{(1)}$, $A_{(2)}$ and $H_{(2)}$, one can easily solve the $\phi_{(3)}$ equation of motion, as can be seen from the Appendix.}

Now, we can find a solution to (\ref{phi1}) by solving the eigen problems
\be \label{EigenTZ}
\left( \!\frac{\pd^2}{\pd t^2} + 3 h \frac{\pd}{\pd t} \!\right) \!\phi_{(1)} = \lambda \,\phi_{(1)} \qquad {\rm and} \qquad e^{2 A_{(0)}} \!\left( \!\frac{\pd^2}{\pd z^2} + 4 A'_{(0)} \frac{\pd}{\pd z} \!\right) \!\phi_{(1)} = \lambda \,\phi_{(1)}
\ee
with some $\lambda = const$. To do that, let us make the ansatz:
\be
\phi_{(1)} = \Phi_1 (t) \,\Phi_2 (z) \,\, .
\ee
Substituting this into the first equation of (\ref{EigenTZ}), we find:
\be \label{Phi1tsol}
\Phi_1 (t) = C_1 \,e^{k_+ t} + C_2 \,e^{k_- t} \,\, ,
\ee
where $C_{1,2}$ are integration constants and
\be \label{kpm}
k_{\pm} = -\frac{3 h}{2} \pm \frac{\sqrt{9 h^2 + 4 \lambda} }{2} \,\, .
\ee
For convenience, below we will take the solution to be
\be \label{Phi1ekt}
\phi_{(1)} (t,z) = e^{kt} \,\Phi (z) \,\, ,
\ee
where $k$ is any of $k_{\pm}$ and we have dropped the index $2$ from the function $\Phi_2 (z)$ for notational simplicity; also, we have absorbed the constant $C_{1,2}$ of (\ref{Phi1tsol}) into the function $\Phi(z)$ in (\ref{Phi1ekt}). 

Now, let us find $\Phi(z)$ by solving the second equation in (\ref{EigenTZ}). Substituting the expression for $A_0 (z)$ from (\ref{ZeroOrSol}), we find that the equation $e^{2 A_0} \!\left( \phi_{(1)}'' + 4 A_0' \phi_{(1)}' \right) \!= \!\lambda \,\phi_{(1)}$ acquires the form:
\be
\frac{7}{3} (z+C)^2 h^2 \phi_{(1)}'' + \frac{28}{3} (z+C) h^2 \phi_{(1)}' - \lambda \phi_{(1)} = 0 \,\,\, .
\ee
The solution of this equation is:
\be \label{PhizC3C4Sol}
\Phi(z) = C_3 (z+C)^{\alpha_+} + C_4 (z+C)^{\alpha_-} \,\,\, ,
\ee
where $C_{3,4}$ are integration constants and
\be \label{alphapm}
\alpha_{\pm} = - \frac{3}{2} \pm \frac{3}{2} \sqrt{1 + \frac{4 \,\lambda}{21 \,h^2}} \,\,\, .
\ee
For convenience, we will take the solution for $\Phi$ to be
\be \label{Phi_z_sol}
\Phi(z) = \tilde{C} \,(z+C)^{\alpha} \,\,\, ,
\ee
where $\alpha$ is any of $\alpha_{\pm}$ and $\tilde{C}$ is a constant.

To summarize, we found that the leading ${\cal O}(\gamma)$ solution for the scalar $\phi$ has the form:
\be \label{phisolTZ}
\phi_{(1)} = \tilde{C} \,e^{kt} \,(z+C)^{\alpha} \,\, ,
\ee
where $k$ and $\alpha$ are given by (\ref{kpm}) and (\ref{alphapm}) respectively. For future use, let us note that if $\lambda$ vanishes, then $k_+ = 0$ and $\alpha_+ = 0$. Hence for $\lambda = 0$ any constant is a solution, as can be seen also directly from (\ref{EigenTZ}).

\subsection{Equations for the metric functions}

Now we turn to the metric warp factors $A(t,z)$ and $H(t,z)$. To begin, let us compute the nonzero Ricci tensor components for the metric (\ref{TimeDepMetricAnz}). The results are:
\bea \label{RicciTDef}
R_{tt} &=& - 3 \ddot{A} - 3 \dot{A} \dot{H} - 3 \ddot{H} - 3 \dot{H}^2 + e^{2A} \left( 4 A'^2 + 3 A' H' + A'' \right) \,\, , \nn \\
R_{x^i x^i} &=& e^{2H} \left( \ddot{A} + 5 \dot{A} \dot{H} + \ddot{H} + 3 \dot{H}^2 + 2 \dot{A}^2 \right) - e^{2A+2H} \left( 4 A'^2 + 7 A'H' + 3 H'^2 \right. \nn \\
&+& \left. A'' + H'' \right) \,\, , \nn \\
R_{zz} &=& - 4 A'' - 4 A'^2 - 6 A' H' - 3 H'' - 3 H'^2 \,\, , \nn \\
R_{tz} &=& - 3 \dot{A}' - 3 \dot{A} H' - 3 \dot{H}' - 3 \dot{H} H' \,\, ,
\eea
where we have again denoted $\dot{} \equiv \frac{\pd}{\pd t}$ and $' \equiv \frac{\pd}{\pd z}$.

Substituting (\ref{RicciTDef}), as well as (\ref{TimeDepScalarAnz}) and (\ref{Expansions})-(\ref{A0H0}), into the field equations on the second line of (\ref{EoM}) and expanding to second order in $\gamma$, we find a set of four metric equations of motion. We drop the order $\gamma^0$ terms since we are expanding around a zeroth order solution, namely (\ref{ZeroOrSol}), and we find at order $\gamma^2$:
\bea \label{E1E2E3E4}
&&E1: \quad - h^2 \!\left( \frac{7}{3} (z+C)^2 A_{(2)}'' + \frac{56}{3} (z+C) A_{(2)}' + 7 (z+C) H_{(2)}' + 6 A_{(2)} \right) \nn \\
&&\hspace*{1.3cm}+ h \left( 3 \dot{A}_{(2)} + 6 \dot{H}_{(2)} \right) + 3 \ddot{A}_{(2)} + 3 \ddot{H}_{(2)} + \frac{1}{2} \dot{\phi}^2_{(1)} = 0 \,\, , \nn \\
&&E2: \quad h^2 \!\left( \frac{7}{3} (z+C)^2 \left[ A_{(2)}'' + H_{(2)}'' \right] + \frac{56}{3} (z+C) A_{(2)}' + \frac{49}{3} (z+C) H_{(2)}' + 6 A_{(2)} \right) \nn \\
&&\hspace*{1.3cm}- h \left( 5 \dot{A}_{(2)} + 6 \dot{H}_{(2)} \right) - \ddot{A}_{(2)} - \ddot{H}_{(2)} = 0 \,\, , \nn \\ 
&&E3: \quad 4 A_{(2)}'' + 3 H_{(2)}'' + \frac{2}{z+C} \left( 4 A_{(2)}' + 3 H_{(2)}' \right) + \frac{1}{2} \phi'^{\,2}_{(1)} = 0 \,\, , \nn \\
&&E4: \quad 3 \dot{A}_{(2)}' + 3 \dot{H}_{(2)}' + 3 \,h H_{(2)}' + \frac{1}{2} \dot{\phi}_{(1)} \phi'_{(1)} = 0 \,\, ,
\eea
where $E1$ comes from the $(tt)$ component in (\ref{EoM}), $E2$ - from the $(x^i x^i)$ one, $E3$ - from the $(zz)$ one and $E4$ from the $(tz)$ one.

The form of equation $E3$ suggests looking for a solution by taking the functions $A_{(2)} (t,z)$ and $H_{(2)} (t,z)$ to have the same time-dependence as $\phi_{(1)}^2 (t,z)$. So let us make the ansatze:
\be \label{AHans}
A_{(2)} (t,z) = e^{2 k t} \hat{A}(z) \qquad {\rm and} \qquad H_{(2)} (t,z) = e^{2kt} \hat{H}(z) + \hat{C}_H \,\, ,
\ee
where $\hat{C}_H = const$. Note that the function $H_{(2)}$ appears in (\ref{E1E2E3E4}) only through its derivatives, which is why we are free to add the arbitrary constant $\hat{C}_H$. Now we can factor out the time dependence. Substituting (\ref{AHans}) and (\ref{Phi1ekt}) into equations $E1-E4$, we find the system:
\bea \label{syst}
&&E1: \quad - h^2 \!\left( \frac{7}{3} (z+C)^2 \hat{A}'' + \frac{56}{3} (z+C) \hat{A}' + 6 \hat{A} + 7 (z+C) \hat{H}' \right) \nn \\
&&\hspace*{1.3cm}6k h \left( \hat{A} + 2 \hat{H} \right) + 12 k^2 \left( \hat{A} + \hat{H} \right) + \frac{k^2}{2} \Phi^2 = 0 \,\, , \nn \\
&&E2: \quad h^2 \!\left( \frac{7}{3} (z+C)^2 (\hat{A}'' + \hat{H}'') + \frac{56}{3} (z+C) \hat{A}' + 6 \hat{A} + \frac{49}{3} (z+C) \hat{H}' \right) \nn \\
&&\hspace*{1.3cm}-2k h \left( 5 \hat{A} + 6 \hat{H} \right) - 4 k^2 \left( \hat{A} + \hat{H} \right) = 0 \,\, , \nn \\
&&E3: \quad 4 \hat{A}'' + 3 \hat{H}'' + \frac{8 \hat{A}' + 6 \hat{H}'}{z+C} + \frac{1}{2} \Phi'^2 = 0 \,\, , \nn \\
&&E4: \quad 6k (\hat{A}' + \hat{H}') + 3 \,h \hat{H}' + \frac{k}{2} \Phi \Phi' = 0 \,\, .
\eea
For future use, let us make the following remark. It is easy to notice that one obtains exactly the same system as (\ref{syst}) if one substitutes 
\be \label{phi1withaddconst}
\phi_{(1)} = C_{\phi} + e^{kt} \,\Phi (z) \qquad {\rm with} \qquad C_{\phi} = const 
\ee
into (\ref{E1E2E3E4}), instead of substituting simply (\ref{Phi1ekt}). The reason is the same as for the presence of the constant $\hat{C}_H$ in $H_{(2)} (t,z)$ above. Namely, the function $\phi_{(1)}$ enters (\ref{E1E2E3E4}) only through its derivatives. 

The system $E1$-$E4$ seems quite involved. However, it can be reduced to a single second order ODE in the following way. First, we can use three of the four equations to solve algebraically for $\hat{H}''$, $\hat{H}'$ and $\hat{H}$ in terms of $\hat{A}$ and its derivatives. Then, substituting these expressions into the fourth equation, we will obtain a linear ODE for $\hat{A}(z)$, which can be solved easily.

To implement the above procedure, let us begin by solving algebraically $E4$ for $\hat{H}'$:
\be \label{Hd}
\hat{H}' = - \frac{k ( 12 \hat{A}' + \Phi \Phi' )}{6\,(2 k + h)} \,\, .
\ee
Substituting this into $E3$, we find:
\be \label{Hdd}
\hat{H}'' = -\frac{4}{3} \hat{A}'' - \frac{1}{6} \Phi'^2 - \frac{8 \hat{A}'}{3(z+C)} + \frac{k}{3(z+C)} \frac{(12 \hat{A}' + \Phi \Phi')}{(2k+h)} \,\, .
\ee
Also, substituting (\ref{Hd}) into $E1$, we obtain: 
\bea \label{Hv}
\hat{H} &=& \frac{ 7 (z+C)^2 h^2 \hat{A}'' }{36 k (h+k)} +  \frac{ 7 (5k + 4 h) (z+C) h^2 \hat{A}' }{18 k (h+k) (2k+h)} + \frac{(h - 2 k ) \hat{A} }{2k} \nn \\
&-& \frac{7 (z+C) h^2 \Phi \Phi'}{72 (k+h) (2k+h)} - \frac{k \Phi^2}{24 (k+h)} \,\, .
\eea
Finally, substituting (\ref{Hd})-(\ref{Hv}) and (\ref{Phi_z_sol}) into $E2$, we end up with the following equation:
\bea \label{A_hat_EOM}
&&28 \,(2 h+k)  \,h^2 (z+C)^2 \hat{A}''(z) + 56 \,(2 h+k) \,h^2 (z+C) \hat{A}'(z) \\
&&+ \,\tilde{C}^2 \left[ 7 h^2 (h+k) \alpha^2 + 14 h^2 k \alpha - 3 k^2 (3 h+k) \right] (z+C)^{2 \alpha} = 0 \nn \,\,\, .
\eea
Its solution is:
\be \label{Ahatsol}
\hat{A} (z) = - \,\tilde{C}^2 \frac{\left[ 7 h^2 (h+k) \alpha^2 + 14 h^2 k \alpha - 3 k^2 (3 h+k) \right]}{56 \,\alpha \,h^2 \,(2 \alpha +1) (2 h + k)} \,(z+C)^{2 \alpha} - \frac{\hat{C}_1}{z+C} + \hat{C}_2 \,\,\, ,
\ee
where $\hat{C}_{1,2}$ are integration constants. Then, using (\ref{Ahatsol}) and (\ref{Phi_z_sol}), we can find $\hat{H}(z)$ from (\ref{Hv}): 
\be \label{Hsol_inc}
\hat{H} (z) = L \,\tilde{C}^2 (z+C)^{2 \alpha} + \frac{2 (h^2 + 3 k^2)}{3 k (h + 2k)} \frac{\hat{C}_1}{(z+C)} + \frac{(h - 2k) \,\hat{C}_2}{2k} \,\,\, ,
\ee
where the constant $L$ is a rather messy expression in terms of $k$, $\alpha$ and $h$; we will not write it down explicitly, since it will not be needed in the following. 

Note, however, that for (\ref{Ahatsol})-(\ref{Hsol_inc}) to be a solution of the system (\ref{syst}), the derivatives of (\ref{Hsol_inc}) have to coincide with the expressions for $\hat{H}'$ and $\hat{H}''$ given by (\ref{Hd}) and (\ref{Hdd}) respectively, possibly up to fixing some of the arbitrary constants. Unfortunately, one can verify that this is not the case. More precisely, the numerical coefficients in front of the $(z+C)^{\alpha}$ terms cannot be matched.\footnote{The quite messy algebraic equations for the constants, that one obtains from equating (\ref{Hd}), (\ref{Hdd}) with their corresponding expressions following from (\ref{Hsol_inc}), do not have a solution.} This suggests that to find a solution, one would need to take $\alpha = 0$ and thus absorb the $(z+C)^{\alpha}$ terms into the constant ones. We will show that this indeed works in the next subsection, where we will follow a route that leads to more manageable (and illuminating) algebraic expressions for the constants.

\subsection{A solution of the coupled system}  

Inspired by the functional form of (\ref{Ahatsol}) and (\ref{Hsol_inc}), we will now look for a solution of the system (\ref{syst}) by starting with the following ansatz: 
\bea \label{Ans_for_sol}
\hat{A} (z) &=& C^A_0 (z+C)^{2 \alpha} + \frac{C^A_1}{z+C} + C^A_2 \,\, , \nn \\
\hat{H} (z) &=& C^H_0 (z+C)^{2 \alpha} + \frac{C^H_1}{z+C} + C^H_2 \,\, ,
\eea
where $C^A_{0,1,2}$ and $C^H_{0,1,2}$ are as yet undetermined constants. Recall also that 
\be \label{Phisolagain}
\Phi (z) = \tilde{C} (z+C)^{\alpha} \,\, ,
\ee
as we saw in Subsection 3.1.

Substituting (\ref{Ans_for_sol})-(\ref{Phisolagain}) in (\ref{syst}), we obtain:
\bea \label{E14}
&&E1: \quad \left\{ \left( 12 k^2 + 6 kh -6h^2 -\frac{28}{3} h^2 \alpha^2 - \frac{98}{3} h^2 \alpha \right) C^A_0 + \frac{k^2 \tilde{C}^2}{2} \right. \nn \\
&&\hspace*{1.3cm} + \left( 12 k^2 +12kh - 14 h^2 \alpha \right) C^H_0 \Bigg\} \, (z+C)^{2 \alpha} \nn \\
&&\hspace*{1.3cm} + \, \frac{\left( 14 h^2 - 6 h^2 + 6 k h + 12 k^2 \right) C^A_1 + \left( 7h^2 + 12 k h + 12 k^2 \right) C^H_1}{z+C} \nn \\
&&\hspace*{1.3cm} + \, 6 \left( 2k^2 + kh - h^2 \right) C^A_2 + 12 k \left( h+k \right) C^H_2 \,= \,0 \,\,\, , \nn \\
&&E2: \quad \left\{ \left( 6 h^2 - 10 k h - 4 k^2 + \frac{28}{3} h^2 \alpha^2 + \frac{98}{3} h^2 \alpha \right) \,C^A_0 \right. \nn \\
&&\hspace*{1.3cm} + \left( \frac{28}{3} h^2 \alpha^2 + 28 h^2 \alpha - 12 k h -4 k^2 \right) C^H_0 \Bigg\} \, (z+C)^{2 \alpha} \nn \\
&&\hspace*{1.3cm} + \, \frac{ \left( 6 h^2 - 10 k h - 4 k^2 - 14 h^2 \right) C^A_1 - \left( 12 k h + 4 k^2 + \frac{35}{3} h^2 \right) C^H_1 }{z+C} \nn \\
&&\hspace*{1.3cm} + \,\left( 6 h^2 - 10 k h - 4 k^2 \right) C^A_2 - \left( 12 k h + 4 k^2 \right) C^H_2 \, = \,0 \,\,\, , \nn \\
&&E3: \quad (2 \alpha + 1) ( 16 C^A_0 + 12 C^H_0 ) + \alpha \,\tilde{C}^2 \,= \,0 \,\,\, , \\
&&E4: \quad \left( \!12 k C^A_0 + 2 (6 k+3 h) C^H_0 + \frac{1}{2} k \tilde{C}^2 \!\right) \!\alpha \,- \,3 \,\frac{2 k C^A_1 + (2 k + h) C^H_1}{(z+C)^{2 \alpha + 1}} \,= \,0 \nn \,\,\, .
\eea
Notice that for $\alpha = - \frac{1}{2}$ equation $E3$ implies that $\tilde{C} = 0$ and thus the scalar field $\phi_{(1)}$ either vanishes or is at most a constant; see (\ref{phi1withaddconst}). Since we are looking for solutions with time-dependent $\phi_{(1)}$, as this is our would-be inflaton, we will take $\alpha \neq - \frac{1}{2}$ from now on. As a consequence, the two terms in $E4$ above have to vanish independently. From the second term we then find:
\be \label{CA1}
C^A_1 = - \,\frac{2k + h}{2k} \,C^H_1 \,\, .
\ee
The vanishing of the first term, together with equation $E3$ in (\ref{E14}), enables us to solve for $C^A_0$ and $C^H_0$, obtaining:
\be \label{AC0CH0}
C^A_0 = \frac{(k - \alpha h) \tilde{C}^2}{8 (2 \alpha + 1)(k+2h)} \qquad , \qquad C^H_0 = - \,\frac{ k (\alpha + 2 ) \tilde{C}^2}{12 (2 \alpha + 1)(k+2h)} \quad .
\ee 

Now, substituting (\ref{CA1}) and (\ref{AC0CH0}) into $E1$ and $E2$ of (\ref{E14}), we have:
\bea \label{E1E2}
&&E1: \quad \frac{\tilde{C}^2 h \left( 14 h^2 \alpha^3 + 49 h^2 \alpha^2 - 30 hk \alpha - 9 k h + 9 h^2 \alpha - 3 k^2 - 6 k^2 \alpha \right) }{12 (2 \alpha + 1) (k+2h)} \,(z+C)^{2 \alpha} \nn \\
&&\hspace*{1.3cm} - \frac{4h^2 (k+h)}{k(z+C)} \,C^H_1 + 12 k (k+h) C^H_2 + 6(2 k^2 + k h - h^2) C^A_2 \, = \, 0 \,\,\, , \nn \\
&&E2: \quad - \frac{(28 h^2 k + 42 h^3) \alpha^3 + (98 h^2 k + 147 h^3) \alpha^2}{36 (2\alpha +1)(k+2h)} \,\tilde{C}^2 (z+C)^{2\alpha} \nn \\
&&\hspace*{1.3cm} -\frac{(27 h^3 - 24 h^2 k - 12 k^3 - 54 k^2 h) \alpha -6 k^3 - 27 h^2 k - 27 k^2 h}{36 (2\alpha +1)(k+2h)} \,\tilde{C}^2 (z+C)^{2\alpha} \nn \\
&&\hspace*{1.3cm} + \frac{4h^2(k+3h)}{3k(z+C)} \,C^H_1 -4 k (k +3h) C^H_2 + 2 (3 h^2 - 5 k h - 2 k^2) C^A_2 \, = \, 0 \,\,\, .
\eea
Recall that we are considering the case $\alpha \neq - \frac{1}{2}$. Hence each of $E1$ and $E2$ in (\ref{E1E2}) implies that
\be \label{CH1}
C^H_1 = 0 \,\, .
\ee
Furthermore, notice that the form of the expressions in $E1$, $E2$ is such that there are two distinct possibilities, namely $\alpha \neq 0$ and $\alpha = 0$. 

Let us consider first $\alpha \neq 0$. In that case, the terms with $(z+C)^{2 \alpha}$ and the constant terms in any of the two equations in (\ref{E1E2}) have to vanish independently of each other. Therefore, each of $E1$ and $E2$ gives rise to two equations (more precisely, algebraic constraints) for the so far arbitrary constants in the problem. The constraints arising from the constant terms of $E1$ and $E2$ are respectively:
\bea \label{Econst}
12 k (k+h) C^H_2 + 6(2 k^2 + k h - h^2) C^A_2 &=& 0 \,\, , \nn \\
-4 k (k +3h) C^H_2 + 2 (3 h^2 - 5 k h - 2 k^2) C^A_2 &=& 0 \,\, .
\eea
From the first equation in (\ref{Econst}), one can readily find:
\be \label{CH2sol}
C^H_2 = \frac{h-2 k}{2 k} \,C^A_2 \,\, .
\ee
It is easy to verify that the second equation in (\ref{Econst}) is identically satisfied upon substituting (\ref{CH2sol}). So we are left with two constraints arising from the $(z+C)^{2\alpha}$ terms in (\ref{E1E2}), namely:
\bea \label{ultconstr}
&&14 h^2 \alpha^3 + 49 h^2 \alpha^2 - 30 hk \alpha - 9 k h + 9 h^2 \alpha - 3 k^2 - 6 k^2 \alpha \, = \, 0 \,\, , \nn \\
&&(28 h^2 k + 42 h^3) \alpha^3 + (98 h^2 k + 147 h^3) \alpha^2 + (27 h^3 - 24 h^2 k - 12 k^3 - 54 k^2 h) \alpha -6 k^3 \nn \\
&&- 27 h^2 k - 27 k^2 h \, = \, 0 \,\, . 
\eea
At first sight, these two equations can be viewed as constraints fixing two of the three constants $h$, $k$ and $\alpha$. However, recall from (\ref{kpm}) and (\ref{alphapm}) that both $k$ an $\alpha$ are defined in terms of $h$ and $\lambda$. So, ultimately, there are only two undetermined constants and two constraints to solve. It would be rather intriguing if we could find a solution with all constants fixed. Unfortunately, however, it turns out that (\ref{ultconstr}) cannot be solved under the assumption that $\alpha \neq 0$. 

Let us explain this in more detail. Solving the constraint on the first line of (\ref{ultconstr}) for $k$, for example, we find:
\be \label{ksol}
k = \frac{- 30 \alpha - 9 \pm \sqrt{336 \alpha^4 + 1344 \alpha^3 + 1704 \alpha^2 + 648 \alpha + 81}}{6 (2 \alpha + 1)} \, h \,\, .
\ee
Now, upon substituting this expression into the second line of (\ref{ultconstr}), we see that the constant $h$ drops out since it appears only as an overall multiplier. Thus, we are left with an equation for $\alpha$ that contains only algebraic operations and square roots. Its solutions for the $+$ sign in (\ref{ksol}) are:
\be \label{al1}
\alpha = 0 \,\,\, , \,\,\, -\frac{7}{4}+\frac{\sqrt{1897}}{28} \,\,\, ,
\ee
and for the $-$ sign are:\,{}\footnote{Note that, if we take the square of this $\alpha$ equation in a way that eliminates the square root, in which case there is no difference between the $+$ and $-$ signs in (\ref{ksol}), we obtain the algebraic equation 
\be
\alpha \left( 784 \,\alpha^7 + 6272 \,\alpha^6 + 18312 \,\alpha^5 + 24920 \,\alpha^4 + 16873 \,\alpha^3 + 5742 \,\alpha^2 + 915 \,\alpha + 54 \right) = 0 \,\,\,\, , \nn
\ee
whose eight solutions are precisely those given in (\ref{al1})-(\ref{al2}) with $\alpha = -\frac{1}{2}$ being a double root.}
\be \label{al2}
\alpha = -2 \,\,\, , \,\,\, - \frac{1}{2} \,\,\, , \,\,\, -\frac{7}{4}-\frac{\sqrt{1897}}{28} \,\,\, , \,\,\, - \frac{3}{4} \pm \frac{\sqrt{273}}{28} \,\,\, .
\ee
However, it turns out that not all of those values lead to solutions of the equations of motion we are studying. The reason is that for most of them the result is not compatible with the relation between $k$ and $\alpha$ that arises from (\ref{kpm}) and (\ref{alphapm}). To take this relation into account, let us solve (\ref{alphapm}) for $\lambda$ and substitute the result in (\ref{kpm}). We obtain: 
\be \label{ksol2}
k = - \frac{3}{2} h \pm \frac{\sqrt{84 \alpha^2 + 252 \alpha + 81}}{6} \,h \,\, .
\ee
Clearly, only values of $\alpha$, for which (\ref{ksol}) and (\ref{ksol2}) give the same value of $k$, are solutions to the full problem. One can check that, among the $\alpha$ values in (\ref{al1}) and (\ref{al2}), the {\it only one} that satisfies this consistency requirement is
\be \label{Alfin}
\alpha = 0 \,\,\, .
\ee

Let us summarize what we have obtained so far. Collecting (\ref{Ans_for_sol}), (\ref{CH1}), (\ref{CA1}), (\ref{AC0CH0}), (\ref{CH2sol}) and (\ref{Alfin}), we have:
\be \label{SolI}
\Phi (z) = \tilde{C} \quad , \quad \hat{A} (z) = C^A_0 + C^A_2 \quad , \quad \hat{H} (z) = C^H_0 + C^H_2
\ee 
with
\be \label{ConstRel}
C^A_0 = \frac{3}{8} \,\tilde{C}^2 \quad , \quad C^H_0 = - \frac{1}{2} \,\tilde{C}^2 \quad , \quad C^H_2 = - \frac{7}{6} \,C^A_2 \quad ,
\ee
where we have used that $\alpha = 0$ implies $k = -3h$ according to any of (\ref{ksol}), (\ref{ksol2}).\footnote{Obviously, those relations also imply that the other solution for $k$, corresponding to vanishing $\alpha$, is $k=0$. However, we discard this solution here due to the division by $k$ in (\ref{CA1}) and (\ref{CH2sol}).} Recall, however, that until now we were considering the $\alpha \neq 0$ case and grouping the various terms in the equations of motion accordingly. So, at first, one might wonder whether we have found a solution with $\alpha = 0$ or just shown that there is no solution with non-vanishing $\alpha$. Of course, it is very easy to check that (\ref{SolI})-(\ref{ConstRel}) do satisfy the equations of motion.

Alternatively, one could start with the following ansatz: 
\bea \label{Ans_for_sol_II}
\Phi (z) &=& \tilde{C} \,\, , \nn \\ 
\hat{A} (z) &=& C_a + \frac{C^a_1}{z+C} \,\, , \nn \\
\hat{H} (z) &=& C_h + \frac{C^h_1}{z+C} \,\, ,
\eea
where $C_a$, $C_h$, $C^a_1$ and $C^h_1$ are arbitrary constants. Then, substituting (\ref{Ans_for_sol_II}) into (\ref{syst}) and performing the same kind of considerations as before, one would easily conclude again that
\be
C^a_1 = 0 \qquad {\rm and} \qquad C^h_1 = 0 \quad ,
\ee
which implies a solution with no $z$-dependence. The only other constraint one would obtain is:
\be \label{constrF}
28 C_a + 24 C_h + \frac{3}{2} \tilde{C}^2 = 0 \,\, .
\ee  
It is easy to check that this is identically satisfied for (\ref{SolI})-(\ref{ConstRel}) upon setting $C_a = C_0^A + C_2^A$ and $C_h = C_0^H + C_2^H$. Note that the $\alpha = 0$ solution, derived by starting from (\ref{Ans_for_sol_II}), is slightly more general as the only relation between the integration constants is given by the constraint (\ref{constrF}).

In conclusion, we have found a solution with $\alpha = 0$. Note that (\ref{alphapm}) then implies that $\lambda = 0$ as well. This in turn gives $k=0$ or $k = - 3h$ according to (\ref{kpm}). Clearly, we are interested in taking  $k=-3h$ in order to have time-dependence in $\phi_{(1)}$. However, recall the discussion around equation (\ref{phi1withaddconst}). Namely, although our considerations in this subsection were based on the assumption that $\phi_{(1)} = e^{kt} \Phi(z)$, nothing changes if we include an additive constant as in (\ref{phi1withaddconst}), since the system we are solving, i.e. (\ref{syst}), remains the same. Furthermore, according to the discussion in subsection 3.1, in the case of vanishing $\lambda$ one can add for free a constant to any $\phi_{(1)}$ solution, as can be seen directly from (\ref{EigenTZ}). Therefore, using (\ref{AHans}) and (\ref{phi1withaddconst}), we finally conclude that our solution has the form:
\be \label{LeadSol}
\phi_{(1)} = C_{\phi} + \tilde{C} e^{-3ht} \,\, , \quad A_{(2)} = C_a e^{-6ht} \,\, , \quad H_{(2)} = \hat{C}_H + C_h e^{-6ht} \,\, ,
\ee
where the constants $\tilde{C}$, $C_a$ and $C_h$ are subject to the constraint (\ref{constrF}), while the constants $C_{\phi}$ and $\hat{C}_H$ are arbitrary.

Before ending this section, we should make one final remark regarding the deformation equations resulting from (\ref{EoM}) upon substitution of the ansatze (\ref{TimeDepScalarAnz}) and (\ref{Expansions}). At order $\gamma$, there is the single equation for $\phi_{(1)}$ that we studied here. However, at order $\gamma^2$, there are in principle two more equations of motion, in addition to the four metric ones we solved. Namely, despite keeping the scalars $p$ and $x$ undeformed, one finds ${\cal O} (\gamma^2)$ contributions in their field equations due to the deformation of the operator $\nabla^2$, induced by the changes in the metric warp factors. However, since the zeroth order solutions $p_0(z)$ and $x_0(z)$ are $t$-independent, while the metric deformations $A_{(2)} (t)$ and $H_{(2)} (t)$ in (\ref{LeadSol}) are $z$-independent, those order $\gamma^2$ terms in the $p$ and $x$ equations vanish identically. One can immediately see this by substituting $\phi$ with $p_0$ or $x_0$ in the expression (\ref{nabfull}), valid for the action of $\nabla^2$ on any scalar, and expanding $A$ and $H$ there to order $\gamma^2$.

\section{Ultra-slow roll inflation}
\setcounter{equation}{0}

In this section we will interpret the solution (\ref{LeadSol}) as a gravitational dual to a field theory in an inflating 4d spacetime, in the vein of the discussion in Section 2. Recall that, in this context, the 5d scalars are viewed as glueballs in the dual gauge theory. Therefore, since the metric part of the solution is supported by a non-trivial 5d scalar, the 4d inflationary expansion is driven by a glueball. Hence we have a putative model of glueball inflation, with the inflaton being the scalar $\phi$. 

Note that, from a 4d perspective, $\phi$ is a composite state in a strongly coupled gauge sector, and not a fundamental scalar. In accordance with that, our inflationary description is {\it not} in terms of an effective (weakly-coupled) action for $\phi$, but instead in terms of directly addressing the expanding 4d spacetime. This is important to underline in order to avoid a possible confusion regarding the computation of various cosmological observables. Namely, unlike in our case, in a wide class of string-theoretic inflationary models, obtained by considering D3 probe branes in various flux backgrounds, one identifies the inflaton with the (radial) position of the D3 brane and the brane worldvolume with the 4d spacetime. One then derives an effective action for the inflaton, starting from the D3 DBI action.\footnote{The literature on this topic is vast and we will not even attempt to cite every relevant work here. Let us just mention, as a relatively recent example, the paper \cite{EHNT} and references therein.} In that context, one would need to compactify (or introduce a cut-off of) the extra dimensions in order to induce dynamical gravity (the Einstein-Hilbert action) on the worldvolume of the D3 brane. This was first shown back in \cite{HV}. In our case, on the other hand, we are not writing an effective action for the inflaton. Instead, by solving the gauge/gravity-duality-derived 5d system of Section 2, we have computed directly the components of the inflating 4d metric. Specifically, we have obtained an explicit expression for the 4d Hubble parameter, as well as the inflaton, as functions of time. As is well-known (see for example \cite{SW, DB}), these functions entirely determine the inflationary slow roll parameters and power spectra.\footnote{Recall that even the scalar power spectrum can be derived just from metric fluctuations upon using comoving gauge; the relevant scalar degree of freedom, then, is the curvature perturbation. Therefore, once we have obtained the usual 4d inflationary background metric, we can directly use the results of the standard 4d computations of cosmological observables.} 

Let us now turn to investigating our model. To do this, we will compute the standard 4d slow roll parameters that follow from (\ref{LeadSol}). Let us begin by recalling that their exact definitions are given in terms of the Hubble parameter and the inflaton as follows (see for example \cite{DB}):\footnote{The more widely used expressions in terms of derivatives of the 4d scalar potential are actually approximate \cite{DB}.}
\be \label{epetaH}
\varepsilon = - \frac{\dot{{\cal H}}}{{\cal H}^2} \qquad {\rm and} \qquad \eta = - \frac{\ddot{\phi}}{{\cal H} \dot{\phi}} \qquad .
\ee
Clearly, to utilize these expressions, we need the functions ${\cal H}$ and $\phi$ corresponding to our solution. Note, however, that the warp factor $A$ also depends on $t$, as is evident from (\ref{LeadSol}). Therefore, we have to first redefine the time coordinate appropriately. 

So let us introduce a new coordinate $\tau$ according to
\be
d \tau = e^{\gamma^2 A_{(2)} (t)} \,dt \,\,\, .
\ee
Since we are working only up to order $\gamma^2$, this, together with (\ref{LeadSol}), implies that:
\be \label{taut}
d \tau = \left( 1 + \gamma^2 C_a e^{-6 h t} \right) dt \qquad \Rightarrow \qquad \tau = t - \gamma^2 \frac{C_a}{6h} \,e^{-6ht} \,\,\, .
\ee
Inverting (\ref{taut}) gives $t = \tau + \frac{W \left( \gamma^2 C_a e^{-6h\tau} \right)}{6h}$\,,
where $W$ is the Lambert $W$ function. Upon expanding this to order $\gamma^2$, we have:
\be \label{tredef}
t = \tau + \frac{C_a e^{- 6 h \tau}}{6h} \,\gamma^2 + {\cal O} (\gamma^4) \,\,\, .
\ee
Then, substituting (\ref{LeadSol}) and (\ref{tredef}) into (\ref{Expansions}), we find that (to order $\gamma^2$) the 5d metric (\ref{TimeDepMetricAnz}) becomes:
\be
ds_5^2 = e^{2A_0(z)} \!\left[ -d\tau^2 + e^{2H(\tau)} d\vec{x}^2 \right] + \,dz^2 \,\,\, ,
\ee
where
\be
H (\tau) = h \,\tau + \gamma^2 \left[ \hat{C}_H + \left( \frac{7}{6} \,C_a + C_h \!\right) \!e^{-6 h \tau} \right] + {\cal O} (\gamma^4) \,\,\, .
\ee
Therefore, applying (\ref{Hubpar}) but now with \,$\dot{} \equiv \frac{d}{d \tau}$\, , \,we obtain:
\be \label{Hres}
{\cal H} = h - 6h \!\left( \frac{7}{6} \,C_a + C_h \!\right) \!e^{-6 h \tau} \gamma^2 + \,{\cal O}(\gamma^4) \,\,\, .
\ee
Also, again using (\ref{LeadSol}) and (\ref{tredef}) in (\ref{Expansions}), one can immediately see that:
\be \label{phires}
\phi = \gamma \left( C_{\phi} + \tilde{C} e^{-3 h \tau} \right) + \,{\cal O} (\gamma^3) \,\,\, .
\ee 

Now we are ready to compute the slow roll parameters. Substituting (\ref{Hres}) and (\ref{phires}) into (\ref{epetaH}), where again $\dot{} \equiv \frac{d}{d \tau}$\,, we find:
\be \label{epf}
\varepsilon = - 36 \!\left( \frac{7}{6} \,C_a + C_h \!\right) \!e^{-6h\tau} \gamma^2 + {\cal O}(\gamma^4)
\ee
and
\be \label{etaf}
\eta = 3 + 18 \!\left( \frac{7}{6} \,C_a + C_h \!\right) \!e^{-6h\tau} \gamma^2 + {\cal O} (\gamma^4) \,\,\, .
\ee
Note that we have to take $\frac{7}{6} C_a + C_h < 0$ in order to ensure $\varepsilon > 0$, as required by the weak energy condition. Then, one can see that (\ref{Hres}) implies $\dot{{\cal H}} < 0$ as should be the case in an expanding universe. 

We can see from (\ref{epf}) and (\ref{etaf}) that, as time increases, $\varepsilon$ tends to zero, whereas $\eta$ tends to $3$. In fact, given that $\gamma <\!\!< 1$, we have in general $\varepsilon <\!\!< 1$ and $\eta \sim {\cal O} (1)$. Clearly then our solution cannot correspond to slow roll inflation, as the condition $\eta <\!\!< 1$ is violated. However, our parameter values, namely at leading order
\be
\varepsilon <\!\!< 1 \qquad {\rm and} \qquad \eta = 3 \quad ,
\ee
are completely consistent with a so called ``ultra-slow roll" regime. This kind of inflationary expansion was first considered in \cite{TW}. It results from an extremely flat region in the 4d inflaton potential, which is the reason for its name despite the breakdown of the slow roll approximation. The power spectrum this ultra-slow roll expansion produces was investigated in depth in \cite{WK}, where it was shown that the scalar spectral index is $n_s = 1$ as required for agreement with observations. Note that our result for the form of the inflaton $\phi$ in (\ref{LeadSol}) is in complete agreement with eq. (135) of \cite{WK}.

However, such a regime is unstable and cannot last for more than a few e-folds. This was already noted in \cite{TW}, based on the observation that the curvature perturbation grows rapidly and thus its backreaction would become significant rather fast. Even without repeating these technical computations, one can see an indication of a rapid exit from the ultra-slow roll regime in the fact that the time-dependence of the inflaton disappears exponentially fast. To that effect, recall that the number of e-folds $N$ is determined by $dN = {\cal H} d \tau$. Therefore, to leading order, (\ref{Hres}) implies that $N = h \tau$. Hence, the inflaton (\ref{phires}) behaves as $\phi (N) = const_1 + const_2 \,e^{-3N}$, which means that $\phi$ approaches constant very fast with the increase of the number of e-folds.\footnote{Another approach, utilized in \cite{MMS}, to showing that this kind of inflationary regime is short-lived, is to compute numerically the exact solution and thus show that it starts deviating from the result of the ultra-slow roll approximation after only a few e-folds. It would be rather interesting to perform such a calculation for our case. However, that seems quite demanding technically and so we leave it for the future.} Given the instability of the ultra-slow roll regime, it is clear that it cannot by itself provide a complete inflationary model. Nevertheless, an ultra-slow roll transient stage, preceding ordinary slow roll, could provide an explanation for the CMB power spectrum anomaly at low $l$ (i.e., multipole moment) values, as discussed recently in \cite{CK}. Hence this class of models is of considerable interest for Early Universe cosmology.

In concluding this section, let us make the following remark. Ultra-slow roll is, in fact, a special case of a broader class of inflationary models, in which the slow roll approximation is violated. They encompass fast roll \cite{AL} and constant rate of roll \cite{MMS,MSY} models. One can define a series of slow roll parameters:
\be
\varepsilon_1 = - \frac{\dot{{\cal H}}}{{\cal H}^2} \qquad {\rm and} \qquad \varepsilon_{n+1} = \frac{\dot{\varepsilon}_n}{{\cal H}\varepsilon_n} \quad ,
\ee
where obviously $\varepsilon_1 \equiv \varepsilon$. The work \cite{MSY} found analytical constant-roll solutions, for which the odd parameters $\varepsilon_{2n+1}$ are small, and tending to zero as time increases, while the even parameters $\varepsilon_{2n}$ are finite and of order 1. Interestingly, although our solution is different, one can verify that it has the same property. Namely, the odd slow roll parameters are $\varepsilon_{2n+1} <\!\!< 1$, whereas the even ones are $\varepsilon_{2n} \sim {\cal O} (1)$. In particular, we have that:
\bea
\varepsilon_2 &=& - 6 - 108 \left( \frac{7}{6} \,C_a + C_h \!\right) e^{-6 h \tau} \,\gamma^2 + {\cal O} (\gamma^4) \,\,\,\, , \nn \\
\varepsilon_3 &=& - 108 \left( \frac{7}{6} \,C_a + C_h \!\right) e^{-6 h \tau} \,\gamma^2 + {\cal O} (\gamma^4) \,\,\,\, , \nn \\
\varepsilon_4 &=& -6 - 84 \left( \frac{7}{6} \,C_a + C_h \!\right) e^{-6 h \tau} \,\gamma^2 + {\cal O} (\gamma^4) \,\,\,\, , \nn \\
\varepsilon_5 &=& - 84 \left( \frac{7}{6} \,C_a + C_h \!\right) e^{-6 h \tau} \,\gamma^2 + {\cal O} (\gamma^4) \quad {\rm etc.} \,\,\,\, .
\eea
Hence, at large $\tau$, the slow roll parameters of our solution behave as:
\be
\varepsilon_{2n+1} \rightarrow 0 \qquad {\rm and} \qquad \varepsilon_{2n} \rightarrow - 6 \quad .
\ee
It is worth exploring whether this similarity with the exact solutions of \cite{MSY} has a deeper meaning.

\section{Discussion} 

We studied time-dependent deformations of the solutions of \cite{ASW,ASW2}. For simplicity, we concentrated on the analytical solution in \cite{ASW} and looked for small (in a certain sense) deviations around it. Using a particular ansatz, we were able to solve the equations of motion for the deviations, to leading order in a certain small parameter. Viewing the resulting background as a gravitational dual to a glueball inflation model, one can compute the inflationary slow roll parameters it leads to. We obtained a result that violates the slow roll approximation, but which is in perfect agreement with a so called {\it ultra-slow roll} expansion. Furthermore, we showed that the form of the inflaton in our solution also matches precisely with the ultra-slow roll expectation. Therefore, we have obtained a dual of ultra-slow roll inflation. 

Despite the breakdown of the slow roll approximation, it was shown in \cite{WK} that an ultra-slow roll regime produces a scale-invariant spectrum of perturbations, as is required for consistency with the observational constraints. However, the ultra-slow roll stage can last only a few e-folds and thus has to be succeeded by a standard slow roll one in order to have the needed amount of inflation. Nevertheless, a transient ultra-slow roll period, occurring at the time of horizon exit of the largest CMB scales, could lead to the observed low $l$ anomaly in the CMB power spectrum; see the discussion in \cite{CK} and references therein. Hence this regime is certainly worth further investigation. 

It should be noted, though, that our original motivation was the search for gravity duals of standard slow-roll glueball inflation. We ended up with a dual of ultra-slow roll by studying a deformation around the analytical solution of \cite{ASW}, which was chosen for technical simplicity. It would be very interesting to explore whether one could find duals of slow roll by looking for deformations around (some of) the numerical solutions in \cite{ASW}. It is also possible that different initial ansatze for the deformations around the analytical solution of \cite{ASW} could lead to different kinds of results, including duals of standard slow-roll glueball inflation models. We hope to investigate this in the future.

\section*{Acknowledgements}

I would like to thank J. Conlon, N. Evans, M. Kulaxizi, A. Parnachev, P. Suranyi, L.C.R. Wijewardhana and L. Pando Zayas for useful discussions. I am also grateful to the Simons Workshop in Mathematics and Physics, Stony Brook 2015, for hospitality during the completion of this work. I have received partial support from the European COST Action MP-1210 and the Bulgarian NSF grant DFNI T02/6.

\appendix

\section{Equation for $\phi_{(3)} (t,z)$}
\setcounter{equation}{0}

The $\nabla^2$ operator for the metric (\ref{TimeDepMetricAnz}), with Christoffel symbols (\ref{Chris_symb}), acts on a scalar $\phi (t,z)$ in the following way:
\be \label{nabfull}
\nabla^2 \phi = - e^{-2A} \left( \ddot{\phi} + 2 \dot{A} \dot{\phi} + 3 \dot{H} \dot{\phi} \right) + \phi'' + 4 A' \phi' + 3 H' \phi' \,\,\, .
\ee
Substituting the expansions (\ref{Expansions}) in (\ref{nabfull}), we see that the leading order is $\gamma$ and at that order the equation $\nabla^2 \phi = 0$ becomes exactly (\ref{phi1}).

At the next order, which is $\gamma^3$, we find the equation:\footnote{Note that the term ${\cal G}^i{}_{jk} (\pd_I \Phi^j) (\pd^I \Phi^k)$ in (\ref{EoM}) does not contribute to the $\phi$ field equation, regardless of the order in $\gamma$, due to (\ref{gabzero}) and the fact that the only nonvanishing ${\cal G}^{\phi}{}_{jk}$ component is ${\cal G}^{\phi}{}_{bb}$\,; \,see, for example, (2.10) of \cite{ASW}.}
\bea \label{phi3}
&&\ddot{\phi}_{(3)} + 3 h \dot{\phi}_{(3)} + \left( 3 \dot{H}_{(2)} \dot{\phi}_{(1)} + 2 \dot{A}_{(2)} \dot{\phi}_{(1)} + 2 A_{(2)} \ddot{\phi}_{(1)} + 6 h A_{(2)} \dot{\phi}_{(1)} \right) = \nn \\
&&= \,e^{2 A_{(0)}} \!\left[ \phi_{(3)}'' + 4 A_{(0)}' \phi_{(3)}' + \left( 4 A_{(2)}' \phi_{(1)}' + 3 H_{(2)}' \phi_{(1)}' \right) \right] \,\, .
\eea
Having determined $\phi_{(1)}$ and $A_{(2)}$, $H_{(2)}$ from the field equations at orders $\gamma$ and $\gamma^2$ respectively (as we do in the main text), one can immediately notice that (\ref{phi3}) is just an equation for $\phi_{(3)} (t,z)$. Since its solution is not needed for our purposes, we will not dwell on it here any further. Despite that, we wrote it out explicitly to illustrate the iterative procedure, via which one could in principle solve the equations of motion to any desired order in $\gamma$.

\end{document}